\documentclass[sigconf, nonacm]{acmart}
\AtBeginDocument{%
  }

\renewcommand\footnotetextcopyrightpermission[1]{} 
\setcopyright{none}                    
\renewcommand\footnotetextcopyrightpermission[1]{} 

\usepackage{graphicx}
\usepackage{subcaption}
\usepackage{float}
\usepackage{fancyhdr}

\title{The Order of Recommendation Matters: Structured Exploration for Improving the Fairness of Content Creators}

\author{Salima Jaoua$^{1}$, Nicolò Pagan$^{1}$, Anikó Hannák$^{1}$, Stefania Ionescu$^{2}$}
\affiliation{
  \vspace{0.3em} 
  \institution{$^{1}$University of Zürich \quad $^{2}$ETH Zürich}\country{}
}

\begin{document}
\begin{abstract}
    Social media platforms provide millions of professional content creators with sustainable incomes. Their income is largely influenced by their number of views and followers, which in turn depends on the platform's recommender system (RS). So, as with regular jobs, it is important to ensure that RSs distribute revenue in a fair way. For example, prior work analyzed whether the creators of the highest-quality content would receive the most followers and income. Results showed this is unlikely to be the case, but did not suggest targeted solutions. In this work, we first use theoretical analysis and simulations on synthetic datasets to understand the system better and find interventions that improve fairness for creators. We find that the use of ordered pairwise comparison overcomes the cold start problem for a new set of items and greatly increases the chance of achieving fair outcomes for all content creators. Importantly, it also maintains user satisfaction. We also test the intervention on the MovieLens dataset and investigate its effectiveness on platforms with interaction histories that are currently unfair for content creators. These experiments reveal that the intervention improves fairness when deployed at early stages of the platform, but the effect decreases as the strength of pre-existing bias increases. Altogether, we find that the ordered pairwise comparison approach might offer a plausible alternative for both new and existing platforms to implement. 
\end{abstract}




\maketitle

\section{INTRODUCTION}
For many professional content creators, visibility and revenue on social media are heavily influenced by recommender system (RS) algorithms. An important issue for RSs, widely tackled by researchers, is the fairness of recommendations which can impact multiple stakeholders. The notion of individual fairness (IF) in RSs extends beyond social media applications to diverse domains, including but not limited to e-commerce, streaming services, and job matching platforms~\cite{IF_1, IF_2}. Although recent advances have been made, current recommendation models struggle to effectively ensure IF without compromising on user satisfaction~\cite{tradeoffUS_if,patro2020fairrec, ranking}.
 
Our work focuses on the IF of creators, i.e, ensuring that creators with similar content quality receive similar audience engagement~\cite{dwork2011fairnessawareness}. This principle promotes equitable compensation based on content quality, incentivizes continued high-quality contributions, and fosters creator trust in platform mechanisms. Prior work on music streaming platforms reveals how music artists perceive fairness on platforms, underlying the need for transparent RSs that should reduce popularity bias~\cite{CCs_interviews}. As such, while user satisfaction remains one of the main goals of platforms, considering creator-IF could also benefit platforms. However, both empirical~\cite{experiment_salganik} and theoretical~\cite{Ionescu2023TheRO} work show that individual fairness is difficult to achieve because early random differences in popularity increase over time through social influence and recommendation mechanisms. This phenomenon, often referred to as the Matthew effect ("rich-get-richer"), persists even when encouraging more diverse recommendations, such as new content~\cite{richgetricher, Ionescu2023TheRO}.
This effect makes popular creators account for most of the platform's engagement, thus reducing the overall diversity of content~\cite{lee2019recommender}.

Following established research, we adopt a multi-stakeholder perspective to analyze and mitigate the long-term impacts of recommender systems~\cite{abdollahpouri2020multistakeholder}. In social media platforms, viewers and creators constitute the primary stakeholders, and popularity bias undermines fairness for both groups~\cite{Ionescu2023TheRO, abdollahpouri2019unfairnesspopularitybiasrecommendation, kowald2019unfairnesspopularitybiasmusic}. Prior work has shown that when a homogeneous user community shares consensus on creator quality, recommender systems exhibit a rich-get-richer dynamic~\cite{article}. Within this simplified setting, we aim to enhance individual fairness for content creators while preserving user satisfaction.
Building upon the agent-based model (ABM) introduced in~\cite{article, Ionescu2023TheRO}, we first establish that perfect individual fairness (IF) is theoretically attainable but computationally prohibitive at scale. We then propose a practical alternative: an ordered pairwise comparison mechanism. By eliciting relative quality judgments through ordered pairwise comparisons, our method presents an effective approach to the new community cold-start problem~\cite{cold_start} by efficiently estimating creator rankings. This method requires only two exploration steps per comparison, thereby leveraging user feedback more effectively~\cite{exploration2021}. We include simulations on synthetic data to evaluate the effectiveness of the proposed intervention in the context of a new community cold-start problem. Finally, we examine its properties and behavior on a real-world dataset.

Our contribution is threefold. First, we propose a novel methodology to promote individual fairness for content creators. Second, we provide both theoretical guarantees and numerical evidence that the proposed intervention improves individual fairness more effectively than exploration-based approaches, without compromising user satisfaction. Importantly, this offers a promising solution to overcome the new community cold-start problem in a fair way. Third, using the widely adopted MovieLens dataset as a benchmark, we demonstrate that the intervention remains effective for established platforms, albeit with diminishing returns.

\section{RELATED WORK}
\label{sec:relatedwork}
\emph{Network Formation}
Network formation in social media platforms has been studied through the lens of preferential attachment~\cite{pa1999, jeong2003measuring}, where nodes acquire connections proportional to their existing degree. \citeauthor{article} \cite{article} instantiate this mechanism in a model where platforms recommend creators with probability proportional to their follower counts, while users strategically follow creators based on content quality. In their seminal paper, they propose a mechanism, validated against data from Twitter and Twitch, where viewers exhibit maximizer behavior~\cite{schwartz2004paradox, schwartz2002maximizing}, searching for the optimal creator to follow. In contrast, \citeauthor{schwartz2002maximizing} also identify satisficer behavior, where viewers accept good enough options based on threshold criteria rather than seeking optimal option. 
Empirical work~\cite{ms_vs_sc_RS} analyzed both behaviors in the context of recommendation and showed many users choose the first recommended option (thus motivating starting with one recommendation per time-step), regardless of their type of behavior (motivating the initial focus on maximizers). 

\citeauthor{Ionescu2023TheRO} \cite{Ionescu2023TheRO} extend this model to investigate fairness for individual creators. Using agent-based simulations, they show that while popularity-based recommendations may be fair in expectation, the probability of achieving fair outcomes for any given creator remains low and varies with the recommendation mechanism. We adopt this modeling framework to study efficient mechanisms for achieving individual fairness. We show that perfect fairness is computationally prohibitive at scale and propose an ordered pairwise comparison mechanism as a practical alternative.

\emph{Fairness in Recommender Systems.} 
Algorithmic bias and fairness in recommendation have gained significant attention across multiple dimensions and stakeholder perspectives \cite{tradeoffUS_if, sonboli2022multisided, burke2017multisidedfairnessrecommendation}. Within this field, researchers often discuss the concepts of group fairness and individual fairness \cite{binns2020apparent, fairness_survey}.  While group fairness metrics aim to ensure equitable outcomes for different demographic groups, individual fairness, first proposed by \citeauthor{dwork2011fairnessawareness}, is defined on the principle that similarly qualified individuals should receive similarly quality outcomes. Surveys of multistakeholder recommendation approaches show the trade-offs between user satisfaction, provider exposure, and platform utility \cite{abdollahpouri2020multistakeholder}. Besides fairness considerations in recommendation outcomes, researchers have also focused on fairness in ranking, showing the fact that the position of items directly influences their exposure and engagement \cite{li2023fairness}. More recently, fairness research shows the importance of addressing dynamic effects. Feedback loops in recommender systems amplify biases over time \cite{mansoury2020feedback, pagan2023classification}.However, while much prior work addresses fairness through on post-hoc debiasing ranking or algorithm modifications, we focus on ensuring individual fairness during the exploration phase via ordered pairwise comparison. At a general level, our work aligns with previous efforts in the recommender systems community to mitigate noise and biases in recommendation algorithms~\cite{guarrasi2024robustrecsys,chen2023bias,ekstrand2022fairness}.

\emph{Cold-start problem and Exploration Strategies.} Another challenge studied in recommender systems and tackled in our work is the cold-start problem, which arises when new content creators or new users join a platform and little to no interaction data is available \cite{cold_start_2}. Traditional research in this field usually focuses on efficient solutions to quickly learn user preferences and item characteristics \cite{cold_start, cold_start_3, cold_start_4}. In addition, we approach this problem by also focusing on fairness, which has received limited attention in existing cold-start literature \cite{zhu2021fairness}. The core challenge lies in balancing exploration (gathering information about new items) with exploitation (leveraging known information to maximize user satisfaction). However, most exploration strategies neglect fairness considerations, despite evidence that initial recommendation decisions can introduce persistent popularity biases~\cite{experiment_salganik, Ionescu2023TheRO}. Existing fairness-aware recommender systems have also focused on fairness in exposure, ensuring that content receive proportional visibility in ranking \cite{singh2018fairness}. For example, envy-freeness form fair division theory ensures that no user prefers another's recommendation allocation \cite{patro2020fairrec}. We focus on fairness of outcomes rather than exposure, as equal visibility does not ensure equal engagement. 
We propose ordered pairwise comparison as a fairness-aware exploration strategy, focusing on fairness of outcomes rather than exposure.  This approach systematically elicits relative quality judgments during the critical early stages of platform development, when creator reputations are most malleable. By directly comparing creators rather than relying solely on absolute engagement metrics, our method mitigates the emergence of popularity-driven inequality while efficiently gathering information about content quality. We demonstrate how this intervention improves creator fairness outcomes during cold-start scenarios without substantially compromising user satisfaction.

\emph{Sequential recommendations.}
Sequential recommendation systems model user preferences as dynamic, treating the sequence of interactions as signals of evolving taste to make recommendation~\cite{Sequential2018, Pi_2019}. Prior work uses temporal feedback to model consumer behavior as sequential. For example, \citeauthor{SIREN2019} captures how users' preferences change in response to the recommended news content \cite{SIREN2019}. \citeauthor{Hodgson2019} develop a Gaussian Process model of product search where learning about one  product update beliefs about similar products \cite{Hodgson2019}. In our work, the model assumes stable user preferences regarding content quality, with temporal dynamics arising from context biases rather than from changing user tastes.



\emph{Other uses of pairwise comparison.} Pairwise comparison proved to be effective for other applications. For example, fine-tuning large language models can be done using pairwise human preferences to train preference models with human feedback\cite{munos2023nash}. In personalized recommendations, pairwise comparison is applied to learn user preferences by asking users to choose between pairs of items rather than provide absolute ratings~\cite{10.14778/2809974.2809992}. Applied to the cold-start problem, pairwise comparison is shown to learn user profiles faster than traditional methods~\cite{Rokach1012}. It is also used to define fairness metrics that assess relative exposure between items~\cite{Beutel2019}. In crowdsourcing, pairwise comparison helps efficiently collect ranking data~\cite{crowdsourcing}. In this work, we argue that a simple pairwise comparison does not guarantee fairness due to asymmetric exposure with a maximizer user behavior, and thus we introduce an ordered method. 


\section{METHODS}
\subsection{Model}
\label{sec:model}
As mentioned in the introduction, chances of achieving individually fair outcomes remain low even in simplified settings where there is one community of users which agree in their evaluation of content creators~\cite{Ionescu2023TheRO, article}. Our paper aims to investigate the causes of unfairness and explore potential interventions in the simplified scenario. As such, we start from the agent-based model of~\citet{Ionescu2023TheRO, article}, which we present below.

The modeled platform consists of regular users (i.e., viewers who solely consume content) and content creators (CCs) forming a bipartite network. The model assumes users have homogeneous preferences (e.g., given by the quality of the content) and thus agree on the relative ordering of CCs. As a result, we can rank CCs by quality, denoting them as $CC_1, CC_2, \dots, CC_n$, with $CC_1$ having the highest quality, and $n$ being the number of CCs on the platform. Note that we do not make any assumptions about the exact value of the quality. It is used purely for ordinal ranking purposes. 

Initially there are no interactions between users and items, thus facing a new community cold-start problem~\cite{Bobadilla2012, Schein2002}. From this state, the platform evolves iteratively. More precisely, each time-step consists of two phases: (a) each user is recommended one content creator, and (b) the user follows the recommended CC if and only if their quality is higher than that of the user's previously followed CCs. It is important to note that the model recommends creators rather than individual content items. This corresponds to a setting where creators are consistent in the quality of the content they create. For one of the results we relax this assumption and introduce noise. We define the popularity-based RS, which recommends CCs with a probability proportional to their current number of followers, i.e, the probability of recommending $CC_i$ at time-step $(t+1)$ is given by $$P^{t+1}(CC_i) = \frac{1+a_i^t}{\sum\limits_{j= 1}^n(1+a_j^t)},$$
where $a_i^t$ corresponds to the number of followers of $CC_i$ at time $t$. As a baseline, we also define the random RS where the recommendation probability is uniformly distributed across all CCs.

\emph{\textbf{Evaluation metrics.}} For evaluation, we adopt a multistakeholder perspective. More precisely, we measure individual fairness for content creators and user satisfaction with respect to the consumed content. Following \citeauthor{Ionescu2023TheRO} \cite{Ionescu2023TheRO}, we say an outcome is $CC_i$-fair if $CC_i$ is at most the $i$th most followed CC. To measure user satisfaction, we compute the average position of the best followed CC. Since $CC_1$ represents the highest quality, this metric measures user dissatisfaction. Intuitively, a low value indicates that users reached, in average, high-quality creators, while a higher value indicates low user satisfaction. 
\begin{definition}[Individual Fairness]
    Let $CC_i$ be the $i^{th}$ best-quality creator and $a^t_i$ the number of followers of $CC_i$ at time $t$. We then say that an outcome $a^t$ is $CC_i$-fair (fair for $CC_i$) if $|\{ j |a^t_j \geq a^t_i \}| \leq i$; i.e., “$CC_i$ is at most the $i^{th}$ most followed creator”.
    We say that an outcome $a^t$ is fair if it's $CC_i$-fair for all $i = 1, \dots, n$.
\end{definition}
The network evolution process, denoted $(A^t)_{t\geq 0}$, is an absorbing Markov Chain, eventually converging to a stable network where no content creator would receive new followers~\cite{Ionescu2023TheRO}.

\subsection{Intervention \& Theoretical results}
Our first goal is to understand how we can improve individual fairness for the setting described above, starting from an empty follower network. As mentioned before, prior work showed that achieving fair outcomes is unlikely even when relying heavily on exploration within the recommender system. To understand why this is the case and investigate potential solutions, we use a theory-driven approach. 

We start from a simple recommendation strategy that guarantees individual fairness throughout the process (first subsection). Although itself not feasible, the proof's insights lead to a feasible alternative: assigning viewers to balanced-sized groups and recommending each group a different ordered pair of creators in the first two rounds of recommendations. We show this strategy still guarantees fairness after the first two time-steps, thus having a good chance of promoting fairness in subsequent rounds (see second subsection). Finally, we demonstrate that noise in viewer's evaluations can be overcome by taking larger viewer group sizes.


\subsubsection*{\textbf{A simple recommendation strategy for perfect IF}}
One strategy for guaranteeing fairness is through perfect symmetry in recommendations. Due to the homogeneity in viewer behavior, we can prove it is sufficient to assign each viewer a permutation of existing content creators and follow this permutation's ordering for providing recommendations. Intuitively, this ensures exposure is envy-free for every pair of creators, not only with respect to the number of times they are recommended as in prior work \cite{singh2018fairness, biega2018equity}, but also with respect to the context in which they are presented. 

\begin{theorem}
    If the platform has $n$ content creators and $k \cdot n!$ users (where $k \in \mathbb{N}$) we can achieve fairness for all CCs by showing a unique permutation to $k$ users in the first $n$ time steps.\footnotemark[2]
\end{theorem}
\begin{proof}
    For simplicity,  we fix $k = 1$. 
    Let $a^t_i$ represent the number of followers of content creator $CC_i$ at time $t$.
 All users are eventually shown $CC_1$ and thus follow $CC_1$ as it has the best quality. Hence, $a^n_1 = n!$. 
More generally, the number of followers of $CC_i$ corresponds to the number of permutations where $CC_i$ is shown before any other creator of higher quality, i.e., $CC_1, \dots, CC_{i-1}$. The fraction of permutations of the set $\{CC_1, \dots, CC_i\}$ that have $CC_i$ placed first is $\frac{1}{i}$. Hence, $ a^n_i = \frac{n!}{i}$.
After $n$ time steps, we reach an absorbing state. Hence, we have $a^t_1 > a^t_2 > \dots > a^t_n$ at $t\geq n$. 
For $k>1$, we generalize the proof by considering that each permutation is presented to $k$ users, and thus the number of followers of each CC is multiplied by $k$.
\end{proof} 
\begin{figure}[htp]
\centering
\includegraphics[width=9cm, trim = 37 450 0 20, clip]{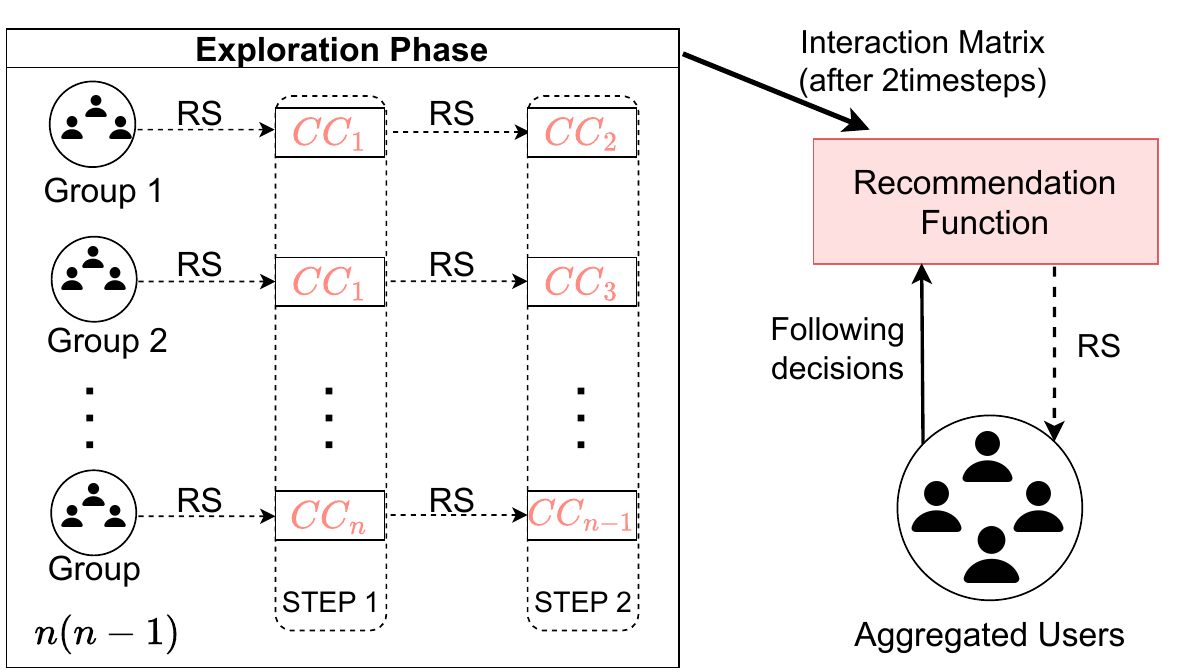}
\caption{Overview of a recommender system using ordered pairwise comparison for early exploration. The exploration phase can be followed by any recommendation algorithm.}
\label{fig:image}
\end{figure}

\subsubsection*{\textbf{Fair exploration of content creators}}
The strategy presented above is not feasible for two main reasons: (1) the number of necessary users grows factorially with the number of CCs, and (2) it requires $n$ time steps of only exploration, leading to long search times and low satisfaction for users. 
However, the analysis revealed that equitable exposure for pairs of similar-quality creators with respect to higher-quality creators is key in promoting fair outcomes.
Using this insight, we can design an efficient and fair short exploration phase (requiring only $2$ time-steps). 

More precisely, we consider the
$n(n - 1)$ ordered pairs of CCs, divide the users into $n(n -1)$ groups of equal sizes, and recommend each user the corresponding pair of creators within the first two time-steps. Later recommendations can follow any algorithm (e.g., random or popularity-based). We illustrate this system in Figure~\ref{fig:image}. When viewers decide who to follow, they signal which creator they find better. By varying the order, we ensure fairness with respect to the chance of being followed.
This introduces our intervention: an ordered pairwise comparison approach to RSs for improving CC-fairness.

\begin{theorem}\label{thm:fairness}
    Suppose the platform has $n$ content creators and $k \cdot n(n-1)$ users, where $k \in \mathbb{N}$. After the ordered pairwise comparison in the first two time-steps, the outcome is IF for all CCs. \footnote{These results can be extended to one community of satisficers and maximizers with equal distributions in the groups and satisficers having the same threshold.}
\end{theorem}

\begin{proof}[Proof~\ref{thm:fairness}]
       At time-step $t =1$, we recommend each content creator to exactly $(n-1)$ groups of $k$ users, meaning  $a_i^1 = k(n-1)$. At time-step $t=2$, users would follow the recommended creator if and only if this second CC has better quality than the first one. For $i = 1, \dots, n$, there are $(n-i)$ pairs in the following form $(j, i)$ where $j > i$. Hence, $ a_i^2 = k(n-1) + k(n-i) = k(2n-i-1)$, which decreases as $i$ increases.
\end{proof} 
Thus, exploration through ordered pairwise comparison leads to a fair state. In theory, we can then turn to exploitation and solely recommend the best creator. However, in practical implementations, users' tastes may evolve through time and may be subject to noise. An exploitation-only RS may be ineffective, as exploration is important to adapt the recommendations to evolving user preferences and content dynamics~\cite{GRAVINO2019371}.
From a fair state, there are high chances of maintaining fairness in the next time-step.  For example, with a popularity-based RS, we can show that the expected follower gain increases with
quality \footnotemark[2]. This suggests ordered pairwise comparison can be effective in combination with other recommendation algorithms.

\subsubsection*{\textbf{Ordered pairwise comparison is effective in noisy settings when increasing group sizes.}}
In practice, noise exists both due to differences in content quality among posts originating from the same creator and due to inconsistencies in user feedback \cite{noise, implicit_cliks}. Thus, larger group sizes for comparing each pair of creators might be needed. Theorem~\ref{thm:groups} specifies the necessary group size for a given level of noise. We model this by assuming every user, when recommended a CC, has a probability $p \in [0,1]$ to follow that CC if they are better than previously followed CCs, and $1-p$ otherwise.


\begin{theorem}\label{thm:groups}
    Suppose the platform has $n$ content creators,  $k \cdot n(n-1)$ users for some $k \in \mathbb{N}$. Let $p \in [0,1]$ be the signal in user behavior. 
    Then, when $k \geq \frac{20(1-p)(3+4p^2)}{p(2p-1)^2}$  then there is at least $95\%$ chance the best quality content creator is the most followed one during a single pairwise comparison.\footnote{We  include the proofs in our appendix.}
\end{theorem}

\section{SIMULATION RESULTS}
\begin{figure*}[htp]
\centering
    \begin{subfigure}{0.3\textwidth}
        \includegraphics[width=\textwidth]{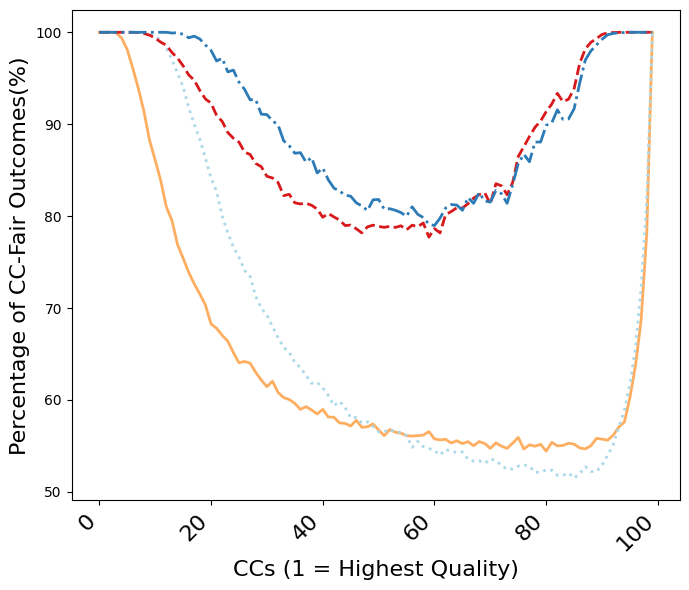}
        \caption{Fairness Distribution by CC Rank } 
        \label{fig:fairness_plot}
    \end{subfigure}\hfill
    \begin{subfigure}{0.3\textwidth}
        \includegraphics[width=\textwidth]{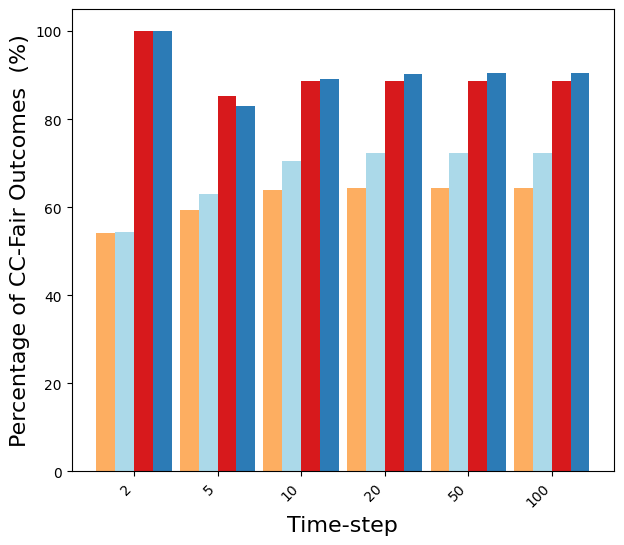}
        \caption{Fairness proportion across time-steps}
        \label{fig:fairness_time}
    \end{subfigure}\hfill
    \begin{subfigure}{0.4\textwidth}
        \includegraphics[width=0.9\textwidth]{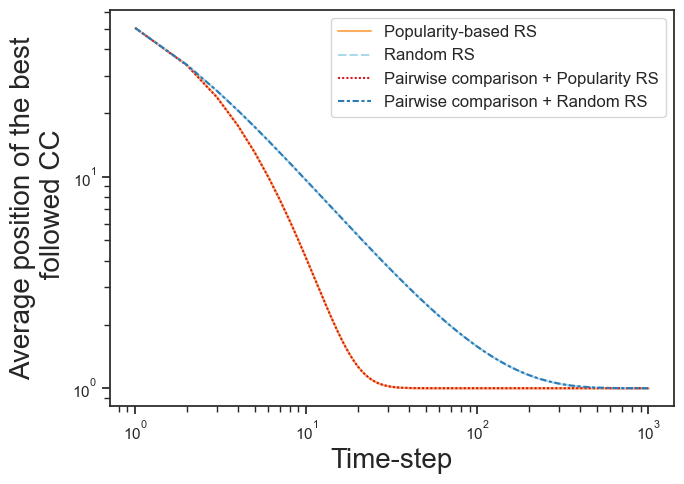}
        \caption{User Dissatisfaction Metrics Across Model Types}
        \label{fig:user_satisfaction}
    \end{subfigure}
\caption{Comparative analysis of recommendation systems with $100$ content creators and $49500$ users: random (Light Blue), popularity-based (Orange), ordered pairwise comparison followed by random  (Dark Blue) and ordered pairwise comparison followed by popularity-based (Red) recommendations. The creators are ranked based on their quality (from the highest quality CC, $CC_1$, to the lowest one, $CC_{100}$).
Plots show (a) the percentage of fair outcomes per content creator across simulations. (b) the percentage of CC-fair outcomes per time-step across simulations. (c) users' dissatisfaction as measured by the position of the best followed CC through time.}
\label{fig:fairness_and_satisfaction}
\end{figure*}
\label{sec:results}
\subsection{Experiments on Synthetic Data}
The results of the previous section show that ordered pairwise comparison improves fairness after the first two rounds. Now, we look at the effects of our intervention on long-term fairness, fairness throughout the process, and viewer satisfaction. 
To perform this analysis, we run simulations on the model presented in Section \ref{sec:model}. As such, the intervention's efficiency is tested for the new community cold start problem with users of homogeneous taste. We apply the intervention to two recommendation systems: the popularity-based RS presented in the previous section, and a random RS representing pure exploration.
For each setting, we take $20,000$ different random seeds and run the simulation for $1000$ time steps. All code and synthetic data is available in our GitHub repository\footnote{Anonymous link: \href{https://github.com/authoranonymous396/Recommend-Me-First}{https://github.com/authoranonymous396/Recommend-Me-First}}. 
We find that ordered pairwise comparison is more effective in improving individual fairness than adding exploration and does not harm user satisfaction. As such, our results suggest it is a promising solution to overcome the new community cold-start problem in a fair way when users have similar tastes.  For example, the ordered pairwise comparison could be implemented in communities where quality is more homogeneous, such as platforms selling construction products, restaurants of a specific cuisine type, or instructional videos. 

\subsubsection*{\textbf{Ordered pairwise comparison improves long-term fairness.}}
Figure \ref{fig:fairness_plot} shows that our intervention ensures each content creator achieves individual fairness in at least $80\%$ of simulation runs, compared to minimum rates of only $57\%$ and $58\%$ for popularity-based and random RS, respectively. The largest difference is observed for mid-range quality CCs, where fair outcomes occur approximately $30\%$  more often. This shows that in the long run, using ordered pairwise comparison greatly increases the chances of achieving fair outcomes for creators.

\subsubsection*{\textbf{Ordered pairwise comparison improves fairness throughout the process}} In realistic settings, reaching convergence (i.e., a state where everybody finds their most preferred content creator) takes a lot of time. Moreover, based on initial performance signals (e.g., followers, engagement), creators of high-quality content can decide to drop out of the platform~\cite{10.1145/3503624}. Therefore, it is especially important to have good levels of fairness early on. Figure~\ref{fig:fairness_time} shows that using pairwise comparison is especially efficient at doing so. More precisely, pairwise comparison leads to perfect fairness for all CCs (100\%) at time-step $t=2$ (result which aligns with Theorem~\ref{thm:fairness}). This is largely maintained in subsequent timesteps (more than $80\%$ at time-step $5$). In contrast, both Popularity-RS and Random RS without the intervention start at only around $55\%$ CC-fair, reaching respectively $65\%, 70\%$ by time-step $100$. This confirms that the ordered pairwise comparison intervention leads quickly to higher fairness and maintains it throughout the process. 

\subsubsection*{\textbf{Ordered pairwise comparison maintains user satisfaction}} Finally, we consider the impact of our intervention on users. This is important because the level of user satisfaction affects not only the happiness of viewers, but also the revenue of the platform, which can impact the platform's willingness to implement the intervention. 
Figure \ref{fig:user_satisfaction} shows that our approach does not harm user satisfaction, measured by the average position (in the quality-ranking scale) of the best content creator followed by the users. 
Unlike exploration-enhancing interventions (such as transitioning to random recommendations), adding the ordered pairwise comparison does not affect the quality of recommendations, hence the users' satisfaction.

\subsection{Experiments on Real-World Data}

The simulation results show that pairwise comparison does not lower user satisfaction and is more effective at improving individual fairness than adding uniform random exploration. 

Another question is whether our intervention can help already running platforms to reduce existing popularity biases~\cite{longtail_2008, Nikolov_2018}. To evaluate this, we conduct experiments using real-world data from the MovieLens dataset~\cite{movielensdataset}. MovieLens is a widely-used benchmark dataset for evaluating RSs, containing  movies, users, movie descriptions, user ratings and timestamps recording when each rating occurred. While MovieLens is well-suited for evaluating prediction accuracy of rating scores, this dataset (like any real-world recommender system dataset) is less suited for long-term fairness evaluation. In particular, the observed ratings reflect the biases encoded in the recommendation algorithm . We lack of information about which movie was recommended to users, preventing us from controlling the influence of the deployed RS. We also lack counterfactual interaction data:  we don't know how users would have engaged under other recommendation. Finally, the dataset is very sparse making the application of our intervention impossible. To address these limitations and enable comparison with our theoretical results, we preprocess the data in several steps described below. 

\subsubsection*{\textbf{Data Processing}} 
We carry out the following three steps. 
\begin{itemize}
    \item[(1)] The pairwise comparison intervention requires binary information, referring to follow or not in our model. To apply this to the data set, we transform ratings using personalized thresholds. Each user's mean rating serves as their threshold: interactions with movies where the rating exceeds the user's mean are assigned $1$ (followed), and $0$ (not followed) otherwise.
    \item[(2)] We apply the ordered pairwise comparison at different moments in the platform evolution. The timestamps allow us to reconstruct the evolution of the interaction matrix and the movie popularity at different stages:  $R^t \in \mathbb{R}^{m \times n },$ where $m$ and $n$ correspond to the number of users and movies, respectively.
    \item[(3)] As we lack data about the decision-making of users under different recommendation, we maintain the maximizer decision-making from our theoretical model: users follow a movie only if its quality is higher than previously followed movies. 
    For the evaluation of quality, we use the IMDb's dataset and ranking formula  \cite{imdb_database,imdb_ratings}\footnote{We note that the popularity distribution from MovieLens ratings does not align with the quality ranking obtained from IMDb, creating a realistic scenario for existing biases.}.
\end{itemize}


To ensure a realistic and computationally feasible setup, we select $100$ movies from the ``film-noir'' genre and $49,500$ users who rated them.

\begin{figure}[htp]
\centering
\includegraphics[width=0.5\textwidth]{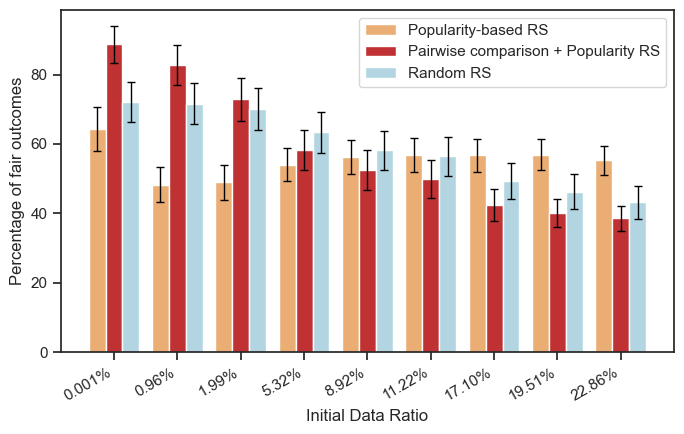}
\caption{Percentage of simulated outcomes that are fair for different recommendation systems - random(Light Blue), popularity-based (Orange), ordered pairwise comparison followed by popularity-based recommendations (Red) - applied starting from different initial ratios of MovieLens interaction matrix evolution. Bars show $95\%$ confidence intervals. The x-axis shows the proportion of MovieLens interactions used to initialize the system.   }
\label{fig:ratio_evolution}
\end{figure}
\begin{figure}
\centering
    \begin{subfigure}{0.9\textwidth}
        \includegraphics[width=0.5\textwidth]{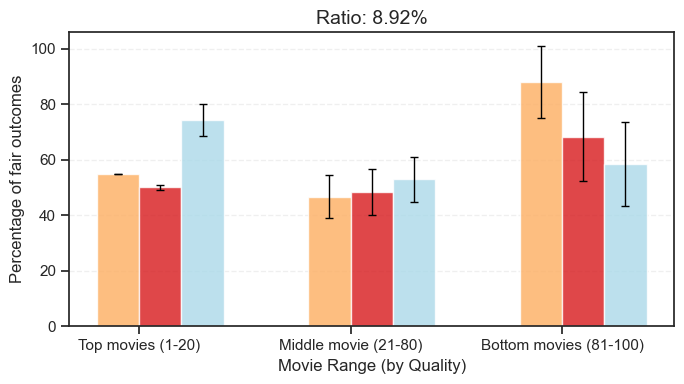}
        \caption{\raggedright Fairness proportion on different Movie ranges}  
        \label{fig:fairness_range}
    \end{subfigure}\vfill
    \begin{subfigure}{0.9\textwidth}
        \includegraphics[width=0.5\textwidth]{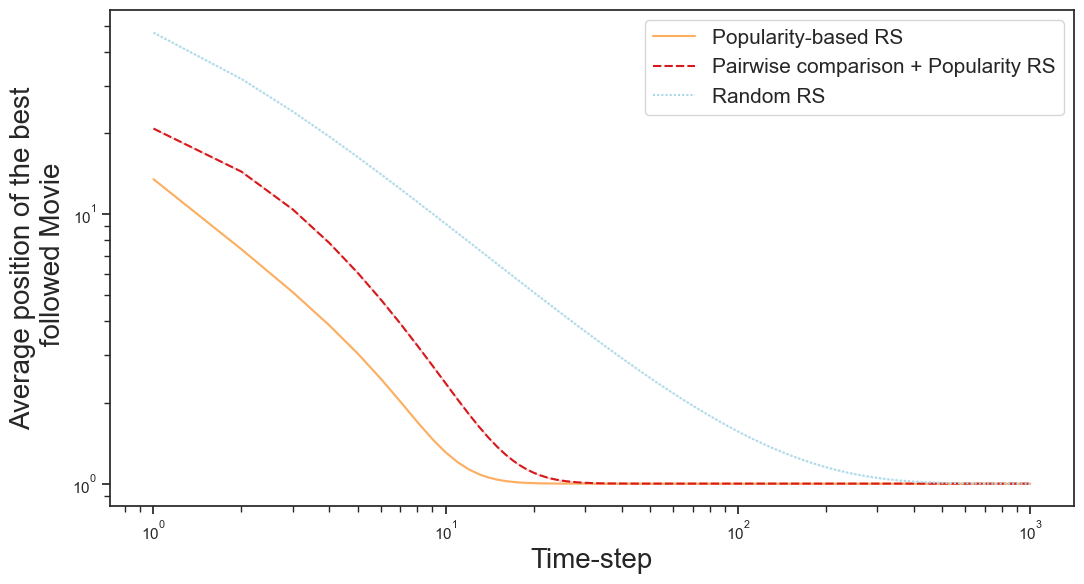}
        \caption{\raggedright User Dissatisfaction Metric}
        \label{fig:user_satisfaction_5}
    \end{subfigure}
\caption{The plots show the results of experiments starting with $8.92\%$ of the initial matrix. (a) Shows the fairness proportion across three different movies groups, representing three ranges of quality rank. Error bars represent $95\%$ confidence intervals. (b) shows the users' dissatisfaction across the $3$ different recommendation systems. Note that we exclude the first two exploration timesteps of the pairwise comparison.  }
\label{fig:ratio532}
\end{figure}
\subsubsection*{\textbf{Ordered pairwise comparison improves fairness for newer platforms}}
As previously anticipated, the real-world dataset allows us to study the effect of the intervention when applied to an already running platform. 
We do so by taking sequential snapshots of the interaction matrix (based on historical observations) and using them as a starting point for simulating our agent-based model under different recommender systems and interventions. For each initial interaction matrix and each recommender system, we run 1,000 simulations.
Figure~\ref{fig:ratio_evolution} reports the results of the average percentage of fair outcomes
across all content creators and simulations as a function of the percentage of initial data taken from real historical interactions.
The results for the initial data ratio of approximately $0\%$ (first columns to the left) are expectedly aligned with the theoretical results: the intervention achieves fairness gains that are comparable to those in the synthetic data simulations, where the interaction matrix is empty (see Figure~\ref{fig:fairness_time} towards convergence).
Furthermore, even though our intervention is calibrated to prevent/reduce unfairness for an initially empty platform, as long as the interaction history used is sufficiently small
(initial data ratio $<5\%$), our intervention still outperforms both the popularity-based and the Random RS, reaching an average of more than $70\%$ of fair outcomes for all creators.
However, the effectiveness of the intervention reduces as initial biases (due to the historical data) accumulate. 
In particular, when $5.32\%$ of the historical data present in the MovieLens dataset are used as initial conditions, all three recommendation algorithms have a similar chance of achieving fair outcomes. For even higher rates of initial data used, fairness outcomes of the proposed intervention progressively reduce to $40\%$ of times (together with the outcomes of the fully exploratory Random RS), while popularity-based RS outperforms the other methods.
Interestingly, when using $12\%$ of the initial data or more, random recommendations achieve lower fairness levels compared to popularity-based recommendations, indicating that full exploration degrades fairness.


To further investigate this effect, we split movies into three categories based on the quality ranking range (top, middle, and bottom quality), and we study the impact of the intervention for the initial interaction matrix built from the first $8.92\%$ of the MovieLens data. 
Figure \ref{fig:fairness_range} shows the fairness proportion for each quality category. Results indicate that bottom-quality movies are 
those affected the most, with the percentage of fair outcomes dropping up to $20\%$ below that of popularity-based RS.
However, for top-quality movies, the intervention performs only slightly slower than popularity-based recommendations. Both recommendation algorithms show very tight error bars, showing that the outcomes are stables. This suggests that  movies have sufficient historical data and established popularity that makes their success deterministic with popularity-based probabilities. Additionally, it suggests that the ordered pairwise comparison does not provide sufficient exposure to overcome these biases. For middle-quality movies, the intervention performs $55\%$ fairness, which is comparable to random RS and popularity-based RS. The results on top- and bottom-quality provide a clear limitation of our intervention when applied to users with pre-existing histories: ordered pairwise comparison can introduce additional unfairness for some movies. Notably, lower-quality movies are most likely to experience this negative impact. We discuss this limitation in more detail at the end of the section.



Finally, we also report the effect of starting from a non-empty matrix on the satisfaction of users: Figure \ref{fig:user_satisfaction_5} shows 
that, even for already running platforms, our approach only marginally affects the rapid convergence that popularity-based recommendations achieve. 

\subsubsection*{\textbf{There exist states where no recommender system can improve fairness.}}
To further investigate why, for initially non-empty interaction
matrices, our intervention has a lower fairness outcomes, we first note that unfairness can be irreversible under maximizer behavior,  and no intervention, including the ordered pairwise comparison, can overcome this. For example, take the case where all users follow $CC_1$ and $CC_3$. This state is unfair as the third movie has more engagement than the second-highest quality movie, $CC_2$. As all users have already converged, no recommendation function can change the state, thus improving fairness. More generally, if we reached an unfair state at convergence, it would remain unfair. This means that once a platform has operated for some time based on misaligned popularity, the structural constraints of maximizer behavior prevent any recommender system from redistributing attention according to quality. Thus, we argue that fairness interventions must be deployed early before such absorbing states emerge. 

\subsubsection*{\textbf{Extending ordered pairwise comparison to be envy-free for creators.}}
Our intervention was designed specifically for new platforms where users lack prior engagement histories. The theoretical principles underlying our approach, particularly the assumption of balanced user populations across comparison groups, hold precisely in this cold-start scenario. For established platforms where users have existing engagement histories, these foundational assumptions no longer apply, necessitating adaptations to the method.

To illustrate why user history matters, consider the following scenario: Let $a_{.1}^t = 2k(n-1), a_{.2}^t = 1, a_{.3}^t = 0$ represent the follower counts of the top-quality, second-quality, and third-quality content creators. If $CC_2$ is shown predominantly to users who have already converged to higher-quality content during exploration, while $CC_3$ is shown to users without such histories, the intervention may inadvertently produce an unfair state where $a_{.3}^{t+2} > a_{.2}^{t+2}$. 

This observation points to a promising direction for extending our approach: applying ordered pairwise comparison to platforms with user engagement histories requires group assignment strategies that account for users' previous (historical) interaction patterns. One natural extension would balance comparison groups based on users' engagement histories (i.e., the last content they positively engaged with in our simplified model). Such an approach would create group assignments that are envy-free for creators, conceptually similar to the user-focused envy-freeness in~\cite{patro2020fairrec}. However, rather than ensuring envy-freeness for users regarding their recommendations, we would structure groups such that no creator prefers another creator's assigned comparison groups over their own.
\section{CONCLUSION}
 \label{sec:discussion}
Building on previous work on individual fairness for content creators, we propose an ordered pairwise-comparison approach for improving fairness on social media platforms. Using an agent-based model, a very useful methodology to reveal important research questions~\cite{gilbert2005simulation} that combines learnings from psychology, complex networks, and recommender systems, allowed us to understand the causal link between cross-session behavior and resulting unfairness, ultimately leading to the proposed intervention. While our theoretical results show that perfect balance in exposure can guarantee fairness, practical implementations must trade this off with feasibility and user satisfaction. Ordered pairwise comparison does precisely this: it ensures balance for the first rounds of recommending new items.

Our theoretical analysis proves that this intervention, when applied to a new community cold-start problem, guarantees individual fairness after the first two rounds. Our simulation results show that our intervention greatly improves fairness both in the short- and long-term. Importantly, using it does not lower user satisfaction.  While our theoretical results only hold in the case of an empty interaction matrix, our simulation with real-world data showed that fairness is also improved with some historical data. Notably, the intervention loses efficacy the more historical data are integrated. This observation has two important and practical implications. Platforms should implement fairness-aware strategies as early as possible, as our results also reveal that correcting bias becomes harder as platforms accumulate interaction history, and in some cases, even impossible. Second, for platforms that have already accumulated very high levels of bias, we need an alternative approach that will account for the information in the historical data.


Our work opens several promising avenues for future research. One natural extension is to apply ordered pairwise comparison to new users joining established platforms. Since these users lack interaction histories, the intervention can provide unbiased quality signals without relying on potentially skewed popularity metrics. Given that this exploration requires only a few time steps, the impact on user satisfaction would be minimal. Combined, these users' implicit and explicit feedback could act as a more accurate evaluation of content quality and serve as a valuable recalibration mechanism for platforms. By examining a subset of creators, platforms can assess their current fairness levels by comparing rankings from new users with those derived from existing data, similar to A/B testing methodologies~\cite{QUIN2024}. Divergent rankings would indicate potential bias, and signals from new users could then inform recommendations to the broader user base.

Our findings depend on the assumption that users behave as cross-session maximizers. So far, the differences in decision-making styles have only been tested for within-session behavior and only for isolated applications~\cite{ms_vs_sc_RS}. Our work shows that users behaving as maximizers~\cite{10.14778/2809974.2809992, Rokach1012} could explain existing levels of unfairness and ground targeted solutions. As such, a potential direction for future work is to design experiments to test whether cross-session user behavior indeed aligns with maximizer's behavior. We hypothesize that maximizer tendencies may be particularly pronounced on platforms featuring short-form content and infinite scroll mechanisms (e.g., TikTok, Instagram), where the interface itself may encourage continuous content consumption and evaluation.

 On real platforms, not all users will have the same preferences. As such, an important focus for future work is to adapt the order-pairwise comparison in the multi-community setting. We have begun exploring multi-community settings with heterogeneous user preferences.Our preliminary results show that the intervention maintains its effectiveness when applied to popularity-based recommender systems in these more complex scenarios. When combined with collaborative filtering (CF) algorithms~\cite{cf2019}, we observe improvements in fairness while preserving user satisfaction levels comparable to baseline approaches. We are currently developing methods to better integrate ordered pairwise comparison with CF by using comparison results to identify user communities, which we expect will yield stronger fairness gains. This integration represents an important step toward scaling our approach to real-world platforms with diverse user populations. One particularly promising direction is leveraging pairwise comparison data not only for fairness interventions but also for community detection, creating a unified framework that addresses both challenges simultaneously.

\begin{acks}
We thank Florian Dörfler, Robin Forsberg, Kshitijaa Jaglan, and the members of the Social Computing group at UZH for their valuable
feedback on this work. This work was supported as a part of NCCR Automation, a National Centre of Competence in Research, funded by the Swiss National Science Foundation (grant number 51NF40\_225155).
\end{acks}
\newpage
\bibliographystyle{ACM-Reference-Format}
\bibliography{IF_for_CC}

\appendix
\section{Theoretical Analysis}
\begin{theorem}
    Assume the network is in a fair state (i.e., with the number of followers ordered by the quality of the creators). Then, a popularity-based RS induces expected follower gains increasing with quality. 
\end{theorem}
\begin{proof}
Let $n$ be the number of content creators, and $a^t_i$ represent the number of followers of content creator $CC_i$ at time $t$.  Being at a fair state means that  we have $a_1^t \geq  a_2^t\geq \dots \geq a_n^t $. Recall that the popularity-based function is defined as  the probability of being recommended $CC_i$ to a user $u$ at time $t+1$: 
$$ \mathbb{P}(R^{t+1}(u) = i ) = \frac{1+ a^t_i}{\sum\limits_{j=1}^n(1+a^t_j)}.$$
Note that this probability does not depends on the user and the denominator is common for every content creators. Hence, for simplicity, we denote this probability as $P^{t+1}(i) = \frac{1+a_i^t}{S_t}$, where $S_t \coloneq\sum\limits_{j=1}^n(1+a^t_j) $. Since we are in a fair state, $P^{t+1}(1) \geq P^{t+1}(2) \geq\dots \geq P^{t+1}(n) $. 
For each $i = 1, \dots, n$, let $\Delta_i^{t+1} $ be the number of followers gained at time $t+1$ by $CC_i$ (i.e., $\Delta_i^{t+1}:= a_i^{t+1} - a_i^{t}$) and $G_i$ be the set of users for which $CC_i$ is the  highest-quality creator they follow at time $t$. 

Now, we show that for an arbitrary $i\in \{1, 2, \dots, n-1\}$ the expected follower gains of $CC_i$ is higher than that of $CC_{i+1}$:
\begin{align*}
    \mathbb{E}[\Delta_i^{t+1}] &= \sum_{j = i+1}^{n} \vert G_{j}\vert \cdot P^{t+1}(i), \\
    &= \vert G_{i+1}\vert \cdot P^{t+1}(i) +\sum_{j = i+2}^{n} \vert G_{j}\vert \cdot P^{t+1}(i), \\
    &\geq \vert G_{i+1}\vert \cdot P^{t+1}(i)+\sum_{j = i+2}^{n} \vert G_{j}\vert \cdot P^{t+1}(i+1), \\ 
    &\geq \vert G_{i+1}\vert \cdot P^{t+1}(i)+  \mathbb{E}[\Delta_{i+1}^{t+1}], \\ 
    &\geq \mathbb{E}[\Delta_{i+1}^{t+1}].
\end{align*}
Since the above holds for all $CC_i$, the claim follows.
\end{proof}
\begin{proof}[Proof of Theorem 3.4]
    Let $(i,j)$ be $2$ two content creators (CCs). With loss of generality we can assume that $i$ have higher quality, meaning $i<j$. From our intervention, we show the pair of CCs to $2$ different group of size $k$ in the two different ordering.  We note the random variable corresponding to  number of followers of respectively $i$ and $j$ at time $t$: $X_i^{(t)}$ and $X_j^{(t)}$. Notice that regardless of their quality,  $X_i^{(1)}\sim Bin (k, p)$ and $X_j^{(1)} \sim Bin (k, p)$. Now at time $t_2$, we have $X_i ^{(2)}\sim Bin (k, p) +  Bin (k, p)$ and  $X_j^{(2)}\sim Bin (k, p) + Bin (X_i^{(1)}, 1-p)+ Bin (k- X_i^{(1)}, p)$.  The two last term describe how users follow $j$ in the second group. Users who already followed $i$ will follow $j$ with probability $(1-p)$. Users who didn't follow $i$ will follow $j$ with probability $p$, as $j$ is perceived as the highest quality CC they haven't yet followed. 

    Note $S := X_i ^{(2)}- X_j^{(2)}$, we want to determine the minimum group size $k$ such that $P (S>0) > 0.95$. We have 
    \begin{align*}
        \mathbb{E}[S] &=  \mathbb{E}[X_i ^{(2)}]- \mathbb{E}[X_j^{(2)}] 
        = 2kp - \mathbb{E}[\mathbb{E}[X_j^{(2)} \vert X_i ^{(1)}] ], \\ 
        &= 2kp - ( kp + \mathbb{E}[X_i ^{(1)}](1-p) + (k-\mathbb{E}[X_i ^{(1)}])p),  \\ 
        &= 2kp - (kp + kp(1-p) + (k-kp)p ), \\ 
        &= kp -2kp (1-p)= kp (2p-1).  
\end{align*}
We have $$\mathbb{V}ar[S] = \mathbb{V}ar[X_i ^{(2)}]+ \mathbb{V}ar[X_j^{(2)}] - 2 \mathbb{C}ov(X_i^{(2)},X_j^{(2)}).$$  
\begin{align*}
     \mathbb{V}ar[X_j^{(2)}] &= \mathbb{E}[\mathbb{V}ar[X_j^{(2)} \vert X_i ^{(1)} ]]+  \mathbb{V}ar[\mathbb{E}[X_j^{(2)} \vert X_i ^{(1)} ]], \\
     &= \mathbb{E}[\mathbb{V}ar[Bin (k, p) \vert X_i ^{(1)} ]] \\ 
     &  \ \ \ \ + \mathbb{E}[\mathbb{V}ar[Bin (X_i^{(1)}, 1-p) \vert X_i ^{(1)} ]] \\ 
     &  \ \ \ \ +\mathbb{E}[\mathbb{V}ar[Bin (k- X_i^{(1)}, p) \vert X_i ^{(1)} ]], \\ 
     &  \ \ \ \ +\mathbb{V}ar[kp + X_i^{(1)}(1-p) + (k- X_i^{(1)})p], \\ 
     &= kp(1-p) +    \mathbb{E}[X_i^{(1)}](1-p)p   
     \\ 
     &  \ \ \ \ + (k-\mathbb{E}[X_i^{(1)}])p(1-p) \\ 
     &  \ \ \ \
     +\mathbb{V}ar[2kp + X_i^{(1)}(1-2p)],\\
     &= 2kp(1-p)+(1-2p)^2kp(1-p), \end{align*}
\begin{align*}
    \mathbb{C}ov(X_i^{(2)},X_j^{(2)})&=    \mathbb{C}ov(X_i^{(1)},X_j^{(2)}), \\ 
    &= \mathbb{C}ov(X_i^{(1)},Bin (X_i^{(1)}, 1-p)) \\ 
     &  \ \ \ \ 
     + \mathbb{C}ov(X_i^{(1)},Bin (k-X_i^{(1)}, p)),  \\
    &= (1-p) \mathbb{V}ar[X_i^{(1)}] - p \mathbb{V}ar[X_i^{(1)}],  \\
    &= (1-p)^2 kp - p^2k(1-p)  \\&= kp(1-p)(1-2p). 
\end{align*}
Hence we have,
\begin{align*}
\mathbb{V}ar[S] &= \mathbb{V}ar[X_i ^{(2)}]+ \mathbb{V}ar[X_j^{(2)}] - 2 \mathbb{C}ov(X_i^{(2)},X_j^{(2)}), \\
    &= kp(1-p) ( 4+ (1-2p)^2-2(1-2p)),\\ 
    &= kp(1-p)( 4+1-4p-4p^2-2+4p), \\
    &= kp(1-p) (3+4p^2).
\end{align*}
Hence , to ensure $\mathbb{P}(S \leq 0)\leq 0.05$ we use Chebyshev's inequality:
\begin{align*}
    \mathbb{P}(S \leq 0)= \mathbb{P} \Bigg( \frac{( S-  \mathbb{E}[S])}{\sqrt{\mathbb{V}ar[S]}}  \leq - \frac{\mathbb{E}[S]}{\sqrt{\mathbb{V}ar[S]}}\Bigg) \leq \frac{\mathbb{V}ar[S]}{\mathbb{E}[S]^2}.
\end{align*} 
We need to find $k$ such that 
\begin{align*}
    \frac{\mathbb{V}ar[S]}{\mathbb{E}[S]^2} \leq 0.05 &\iff \frac{kp(1-p) (3+4p^2)}{ k^2p^2 (2p-1)^2} \leq 0.05m \\ 
    &\iff \frac{(1-p)(3+4p^2)}{kp(2p-1)^2} \leq 0.05, \\ 
    &\iff k \geq \frac{20(1-p)(3+4p^2)}{p(2p-1)^2}.
\end{align*}
\end{proof}
\section{Additional Results}
\subsection*{Ordered pairwise comparison improves fairness throughout the process}
\begin{figure*}[ht]
    \centering
    \begin{subfigure}{0.45\textwidth}
        \includegraphics[width=0.8\linewidth]{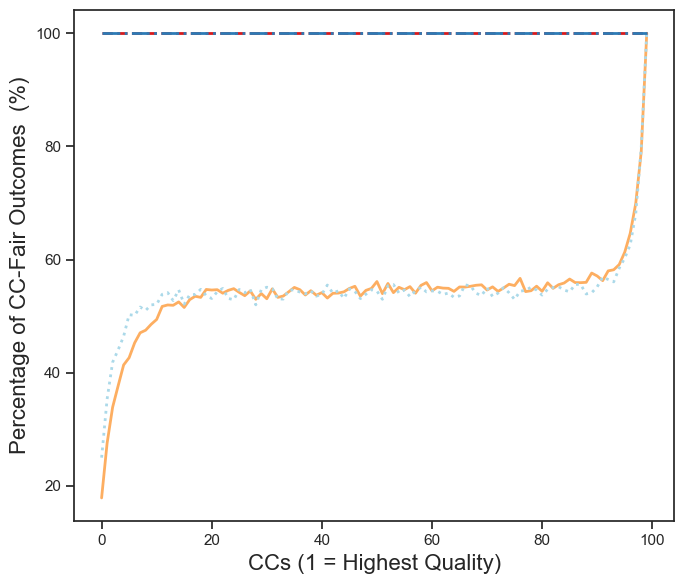}
        \caption{Fairness Distribution by CC Rank at time-step 2}
    \end{subfigure}
    \hfill
    \begin{subfigure}{0.45\textwidth}
        \includegraphics[width=0.8\linewidth]{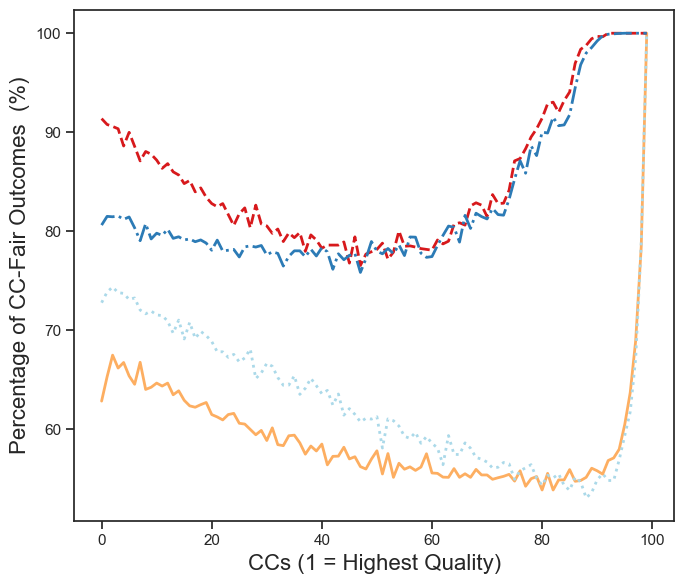}
        \caption{Fairness Distribution by CC Rank at time-step 5}
    \end{subfigure}

    \vspace{0.2em} 

    \begin{subfigure}{0.45\textwidth}
        \includegraphics[width=0.8\linewidth]{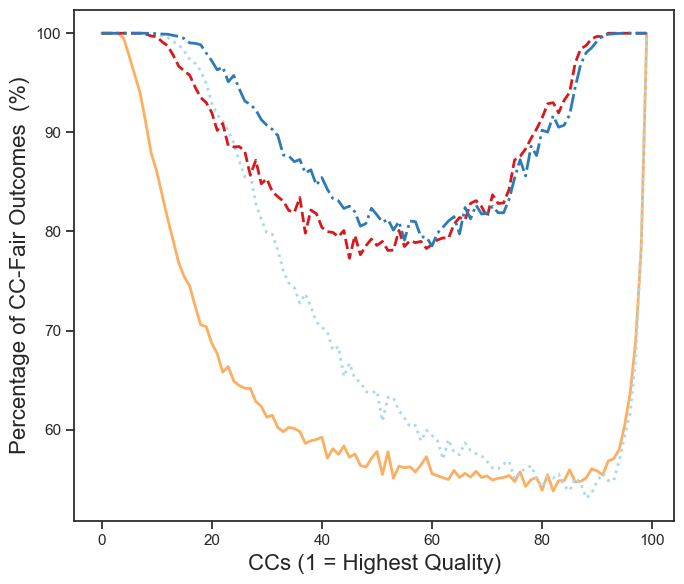}
        \caption{Fairness Distribution by CC Rank at time-step 20}
    \end{subfigure}
    \hfill
    \begin{subfigure}{0.45\textwidth}
        \includegraphics[width=0.8\linewidth]{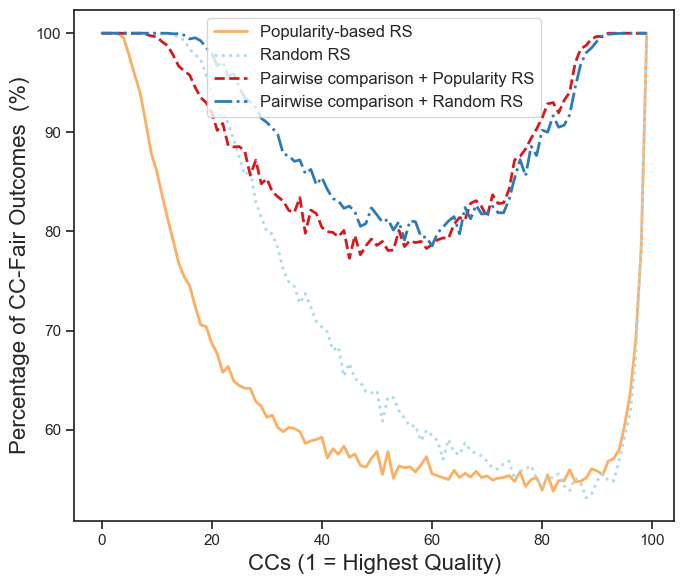}
        \caption{Fairness Distribution by CC Rank at time-step 100}
    \end{subfigure}
\caption{Comparative analysis of recommendation systems with $100$ content creators and $49500$ users: random (Light Blue), popularity-based (Orange), ordered pairwise comparison followed by random  (Dark Blue) and ordered pairwise comparison followed by popularity-based (Red) recommendations. The creators are ranked based on their quality ($CC_1$  highest quality CC to $CC_{100}$ lowest).
Plots show the percentage of fair outcomes per content creator across simulations at different time-steps: (a) $t= 2$(b) $t= 5$(c) $t=20$ (d) $t= 100$.}
\label{fig:fairness_throughout_process_detailed}
\end{figure*}

We plot additional results in Figure \ref{fig:fairness_throughout_process_detailed} to show how fairness evolves from each content creator throughout the recommendation algorithms.



\end{document}